\documentclass[twocolumn,showpacs,showkeys,preprintnumbers,amsmath,amssymb]{revtex4} 

\usepackage{graphicx}
\usepackage{dcolumn}
\usepackage{bm}

\begin{document}

\title{Controlling the polarization eigenstate of a quantum dot exciton with light}

\author{Thomas Belhadj$^1$}
\author{Claire-Marie Simon$^{1,2}$}
\author{Thierry Amand$^1$}
\author{Pierre Renucci$^1$}
\author{Olivier Krebs$^3$}
\author{Aristide Lema\^itre$^3$}
\author{Paul Voisin$^3$}
\author{Xavier Marie$^1$}
\author{Bernhard Urbaszek$^1$}
\email[Corresponding author : ]{urbaszek@insa-toulouse.fr}
\affiliation{%
$^1$Universit\'e de Toulouse, INSA-CNRS-UPS, LPCNO, 135 Av. Rangueil, 31077 Toulouse, France}

\affiliation{%
$^2$Universit\'e de Toulouse, CNRS-UPS, LCAR, IRSAMC, 31062 Toulouse, France}

\affiliation{%
$^3$Laboratoire de Photonique et Nanostructures CNRS, route de Nozay, 91460 Marcoussis, France}

\date{\today}

\begin{abstract}
We demonstrate optical control of the polarization eigenstates of a neutral quantum dot exciton without any external fields.  By varying the excitation power of a circularly polarized laser in micro-photoluminescence experiments on individual InGaAs quantum dots we control the magnitude and direction of an effective internal magnetic field created via optical pumping of nuclear spins. The adjustable nuclear magnetic field allows us to tune the linear and circular polarization degree of the neutral exciton emission. The quantum dot can thus act as a tunable light polarization converter.

\end{abstract}

\maketitle

Semiconductor quantum dots (QDs) are nanometer sized objects that contain typically several thousand atoms resulting in a confinement of electrons in all three spatial directions. The absence of translational motion prolongs the carrier spin lifetimes as compared to bulk (3D) and quantum well (2D) structures \cite{md2008,Kroutvar1,Marie01,bracker08,hanson07}. As a result a large number of schemes for QD spin based q-bit manipulations have been proposed \cite{Henne2009}. After optical excitation, a conduction electron and a valence hole form a neutral exciton X$^0$ in the dot. For the model system of self assembled InGaAs QDs in  GaAs, the anisotropic electron hole Coulomb exchange interaction for the QD symmetry C$_{2v}$ gives rise to a bright X$^0$ doublet of eigenstates $\vert X\rangle$ and $\vert Y\rangle$ polarized along the [$1\bar{1}0$] and [110] crystallographic directions, respectively \cite{Mb2,Bester}. To characterize the strength of the anisotropic Coulomb exchange an effective magnetic field B$_{AEI}$ in the QD plane acting on the exciton pseudo spin can be introduced \cite{Dzhioev97,astakhov06,kowalik2007}.

In analogy to the electron and the proton in a hydrogen atom, the electron in a QD is also interacting with the \emph{magnetic} moment of the nuclear spins of the atoms that form the dot \cite{Abra}. The electron-hole Coulomb exchange interaction cancels out in the ground state of a singly charged exciton as for example the X$^+$ (2 holes + 1 electron) as the holes form a spin singlet \cite{braun2006}.  
Under suitable excitation conditions, the electron polarization created through optical pumping of the X$^+$ exciton can be transferred to the nuclear spins in the dot via the hyperfine interaction even at zero applied magnetic field, giving rise to an effective magnetic field B$_N$ that can in turn stabilize the electron spin \cite{md2008,Lai,dk2007,Pal07}.

\begin{figure}
\includegraphics[width=0.45\textwidth]{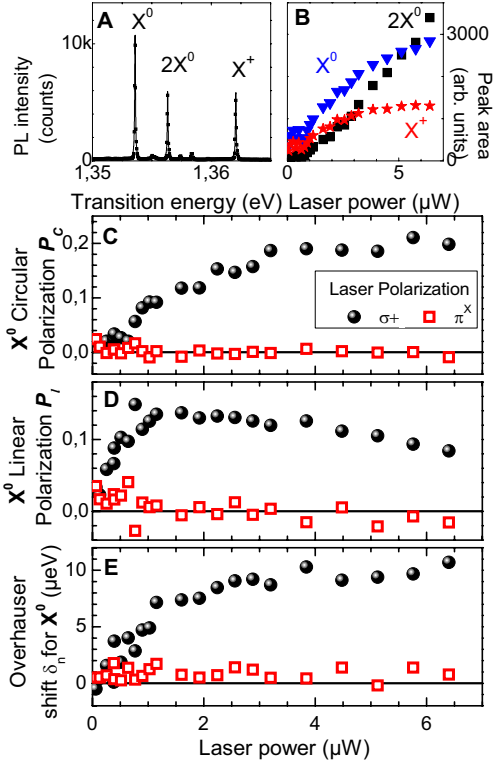}
\caption{\label{fig:fig1} (Color online) QD I. E$_\text{laser}$ = 1.44~eV. (A) PL spectrum at 4K for P=2.5$\mu$W and linear $\pi^X$ excitation and detection. (B) Integrated PL intensity as a function of laser power for the X$^0$ (triangles), X$^+$ (stars) and 2X$^0$ (squares). The circular polarization degree P$_c$ (C), the linear polarization degree P$_{l}$ (D) and the Overhauser shift $\delta_n$ (E) are shown as a function of laser power for X$^0$.  Solid circles (hollow squares) represent $\sigma^+$ ($\pi^X$) laser polarization.}
\end{figure}

In this Letter we demonstrate optical control of the polarization eigenstate of a neutral quantum dot exciton X$^0$ in the absence of any external magnetic or electric field. We show novel effects resulting from the combined effect of the effective nuclear magnetic field B$_N$ and the Coulomb exchange interaction (i.e. B$_{AEI}$) on the electron spin in an InGaAs QD: the control over  B$_N$ via non-resonant optical pumping allows us to orientate the pseudo spin of a neutral exciton and therefore achieve substantial optical orientation, previously only reported for charged excitons. 
As compared to charged excitons, we show that the robust electron spin injection for X$^0$ has the advantage that in the presence of B$_{AEI}$ the quantum dot can act as a \emph{tunable} light polarization converter. The degree of circular to linear polarization conversion can be adjusted through a slight variation in excitation laser power, which could provide a new approach to switching the polarization of QD based single photon emitters \cite{Strauf07}.  We show that the build-up of B$_N$ is possible due to the presence of charged excitons X$^+$ appearing under non-resonant pumping conditions.

The sample consists of: GaAs substrate, 20 nm of GaAlAs, 98 nm GaAs, delta doping Si  10$^9$ cm$^{-2}$, 2 nm GaAs, InGaAs wetting layer (WL) and QDs, 100 nm GaAs, 20 nm of GaAlAs, 5nm GaAs. Although the samples are intentionally n-doped, detailed spectroscopic analysis shows that (residual) p-type doping prevails, leading to the observation of neutral and singly positively charged excitons. The photoluminescence (PL) and PL excitation (PLE) measurements at 4K on individual QDs were carried out with a confocal microscope build around attocube nano-postioners connected to a spectrometer and a charge coupled device (CCD) camera. The signal to noise ratio of 10$^4$ obtained by placing a solid immersion lens on the sample allows to obtain a spectral precision of +/- 1 $\mu$eV for the transition energy by fitting the spectra with Lorentzian lineshapes. The excitation energy E$_\text{laser}$ of a continuous wave Ti-Sapphire laser is varied between 1.38 and 1.48 eV, covering the heavy hole and light hole to electron transitions in the WL \cite{Moskalenko02,Bimberg03} .The circular polarization degree of the QD PL is defined as $P_c = (I^+ - I^-)/ (I^+ + I^-)$, where $I^{+(-)}$ is the $\sigma^{+(-)}$ polarized PL intensity integrated over the spectral domain covering the X$^0$ doublet (X$^+$ singlet) emission. The linear polarization degree is defined as $P_{l} = (I^X - I^Y)/ (I^X + I^Y)$.

\begin{figure}
\includegraphics[width=0.4\textwidth]{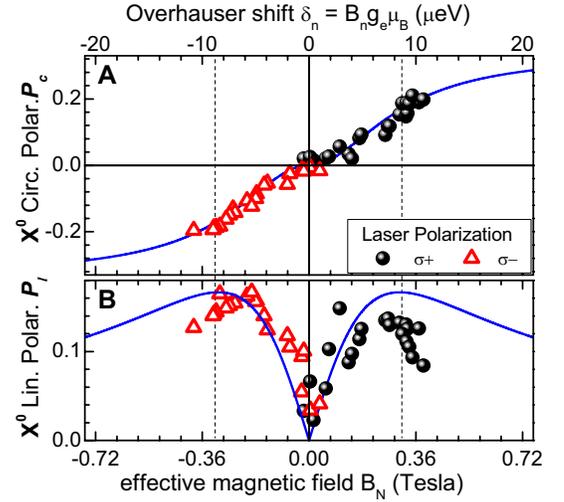}
\caption{\label{fig:fig2} (Color online) QD I. E$_\text{laser}$ = 1.44~eV. $\delta_1 = -9 \mu$eV Circular polarization degree P$_c$ (B) and Linear polarization degree P$_{l}$ (C) for the X$^0$ states as a function of the effective nuclear field B$_N$ for $\sigma^+$ ($\sigma^-$) laser polarization shown as solid circles (hollow triangles) assuming $\vert g_e \vert=0.48$. Solid lines are calculated with equations 2 and 3. Dashed vertical line: measured value of $\vert\delta_1 \vert$. 
 }
\end{figure}

Figure 1.A shows a typical PL spectrum for E$_\text{laser}$=1.44~eV exciting the heavy hole to electron transition in the WL about 90meV above the QD neutral exciton X$^0$ emission. Three transitions are observed: X$^0$, the neutral biexciton 2X$^0$ (2 electrons + 2 holes) which is blue shifted \cite{Bimberg03}, and the positively charged exciton X$^+$. The X$^+$ shows as anticipated no fine structure and a P$_c$ in the order of 50\% under strong pumping with circularly polarized light (not shown) \cite{footxp}. The attribution of the three transitions is in agreement with the power dependence of the PL intensity shown in figure 1B.

Under linearly polarized excitation, the two bright X$^0$ states  $\vert X\rangle = \frac {\vert \Uparrow, \downarrow \rangle + \vert \Downarrow, \uparrow \rangle}{\sqrt{2}}$  and $\vert Y\rangle = \frac {\vert \Uparrow, \downarrow \rangle - \vert \Downarrow, \uparrow \rangle}{i\sqrt{2}}$ are separated in energy by $\delta_1 \equiv E_X - E_Y \simeq - 9 \mu$eV for dot I due to B$_{AEI}$ \cite{Mb2}.
Here  $\Uparrow (\Downarrow)$ stands for the heavy hole pseudo spin up (down)  and $\uparrow (\downarrow)$ for the electron spin up (down) projections onto the \emph{z}-axis, which is also the light propagation axis and the sample growth axis \cite{footLH}. 
The same splitting $\vert\delta_1 \vert$ is found for the 2X$^0$, but as expected with the order of the peaks reversed. 
It is important to note that equal intensities $I^X=I^Y$ of the linearly polarized transitions $\vert X\rangle$ and $\vert Y\rangle$ result in a net PL polarization P$_l=0$ when integrating over both transitions.

A surprising power dependence of P$_c$ and P$_l$ for X$^0$ following circularly polarized excitation in the WL is shown in figure 1C and D. We observe an increase from  P$_c \simeq 0$ up to 22\% when exciting with P$_{exc}$ above 4$\mu$W (figure 1C), so substantial optical orientation has been achieved. 
Even more intriguing, we observe under \emph{circularly} polarized excitation that the \emph{linear} polarization increases abruptly with Laser power from P$_{l} \simeq 0$ to 17\% before gradually decreasing for P$_{exc}>1\mu$W, (figure 1D). These effects do not purely depend on laser power but also polarization, as can be seen in figures 1C and D: exciting with linearly polarized light results in  $P_c \simeq P_l \simeq 0$ for X$^0$ with no dependence on P$_{exc}$ \cite{foot1}.

With only the exchange interaction B$_{AEI}$ present the X$^0$ eigenstates $\vert X\rangle$ and $\vert Y\rangle$ are linearly polarized. Excitation with circularly polarized light should result in beats in the time domain between $\vert\Uparrow,\downarrow\rangle$ and  $\vert\Downarrow,\uparrow\rangle$ as those are not the X$^0$ polarization eigenstates \cite{henne01,steel02}. So assuming (i) an exponential radiative decay for the X$^0$ with a characteristic time $\tau_r = 700 ps $ \cite{Marie01} and (ii) an exciton spin lifetime $\tau_s \gg \tau_r $ the measured P$_c$ in cw PL would be P$_c = P_{c}^0(1+\omega ^2 \tau_r ^2)^{-1}$ with $\hbar\omega=\delta_1$ and P$_{c}^0$ is the P$_{c}$ created in the dot at time t=0 for X$^0$ \cite{astakhov06}.  For the measured $\delta_1\simeq-9 \mu$eV one would only expect P$_c^\text{MAX} \simeq 1\%$, and not 22\% as found in the experiment. Concerning P$_l$, circularly polarized excitation should result in PL with $I^X=I^Y$ and hence P$_l =0$ which is in contradiction to the 17\% measured.

The observed optical orientation and polarization conversion can not be explained without invoking new X$^0$ eigenstates.  We will argue in the following that non-resonant optical pumping has created a dynamic nuclear spin polarization (DNP) that acts on the electron spin like an effective internal magnetic field of several hundred mT along the \emph{z}-axis (see figure 2A and B). The coupling of the nuclear spins to the electron spin via the Fermi-contact interaction (neglected here for hole spins \cite{Abra}) can be expressed as 
\small
\begin{equation}
\label{eq:eqHf1}
H_{HF} = \sum_k^N A_k \left(I_z^k S_z+ \frac{I_-^k S_+ + I_+^k S_- }{ 2 } \right)
\end{equation}
\normalsize 

and $\langle H_{HF} \rangle = A \langle \vec{I} \rangle\vec{S}\equiv g_e \mu_B \vec{B}_N \vec{S}$, where $\vec{I}^k$ and $\vec{S}$ are the spin operator for nucleus $k$ (out of $N\simeq 10^4 - 10^5$) and for the electron spin, respectively. $g_e$ is the longitudinal electron g factor and $I_z^\text{MAX}$ for In, Ga and As is $9/2$, $3/2$ and $3/2$, respectively.
The combined effect of an \emph{external} longitudinal magnetic field and B$_{AEI}$ on the bright exciton doublet are detailed in \cite{Marie01,Mb2,Dzhioev97,kowalik2007,GammonPRL}. Here we simply replace the Zeeman Hamiltonian by $\langle H_{HF} \rangle$ resulting in a Zeeman splitting (called Overhauser shift) purely due to $\vec{B}_N =(0,0,B_N)$ of $\delta_n = g_e \mu_B B_N$.

The presence of a magnetic field component along the \emph{z}-axis will result in \\
(i) a splitting $\sqrt{\delta_1^2 + \delta_n^2}$ of the bright X$^0$ doublet and \\
(ii) new eigenstates $\vert + \rangle = \alpha \vert X \rangle + i\beta \vert Y \rangle$ and \\
$\vert - \rangle = \beta \vert X \rangle - i\alpha \vert Y \rangle$ where $\alpha^2=1$ $(\alpha^2\rightarrow 1/2)$ for B$_N$=0 (B$_N \rightarrow \infty)$ and $\alpha=\alpha(\delta_1,\delta_n)$ and $\beta=\beta(\delta_1,\delta_n)$. Assuming that $\tau_r\gg \Omega^{-1}$ where $\hbar \Omega = \sqrt{\delta_1^2 + \delta_n^2 }$ we find:
\small
\begin{eqnarray} 
\label{equAlpha}
P_c(\delta_N)=4 \alpha^2 \beta^2 P_{c}^0=\delta_n^2 P_{c}^0/(\delta_n^2+\delta_1^2)\\
P_{l}(\delta_N)=2 \alpha\beta (\alpha^2 - \beta^2) P_{c}^0=-\delta_n\delta_1 P_{c}^0/(\delta_n^2+\delta_1^2)
\end{eqnarray}\\
\normalsize 
With our fitting procedure we can extract $\delta_n$ for X$^0$ as a function of Laser power for $\sigma^+$ excitation (see figure 1E) \cite{footFIT}. An identical splitting with opposite sign is found for $\sigma^-$ excitation (not shown) and for linearly polarized excitation we measure $\delta_n\simeq 0$, as $B_N=0$. As the dependence of $\delta_n$ on P$_{exc}$ is non-linear, it is more instructive to plot P$_c$ (figure 2.A)  and P$_l$ (figure 2.B) achieved for X$^0$ as a function of the created field $B_N \propto \delta_n$ (assuming an electron g-factor of $\vert g_e \vert=0.48$ \cite{braun2006}).
The experimental curves in figure 2 are very well reproduced using $\vert P_{c}^0\vert$=33\% as the only fitting parameter in equations 2 and 3. The measured P$_c$ of $\pm$ 22\% for $\sigma \pm$ excitation represents 65\% of the maximum achievable P$_{c}^0$=$\pm$33\% for $\delta_n \to \infty$, although only a low nuclear polarization of roughly 5\% is achieved as $\delta_n\simeq 10 \mu$eV. We demonstrate a wide range of tunability for the circular to linear conversion as we go from P$_l \simeq 0 $  to the theoretical limit of maximum conversion $P_{l}=P_{c}^0/2$ for $ \vert \delta_n \vert = \vert \delta_1\vert $. For $ \vert \delta_n \vert > \vert \delta_1\vert $ P$_{l}$ decreases in both theory and experiment. For P$_l$ not all experimental points are on the theoretical curve and we notice a slight asymmetry between $\sigma^+$ and $\sigma^-$ excitation. Our simple model does not take into account strain induced heavy hole - light hole coupling which results in X$^0$ eigenstates which are already at $B_N=0$ different from $\vert X\rangle$ and $\vert Y\rangle$ \cite{kowalik2007,koudinov}. This could be at the origin of the observed discrepancy.
 
\begin{figure}
\includegraphics[width=0.48\textwidth]{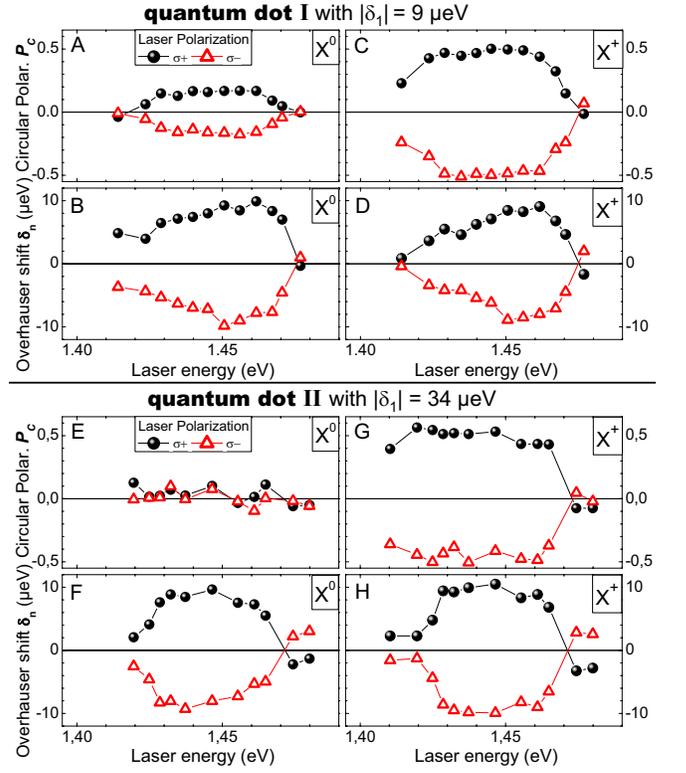}
\caption{\label{fig:fig3} (Color online)  PL detection energy dot I (II) $\simeq$ 1.358~eV ($\simeq$ 1.342~eV). Circular polarization degree P$_c$ and Overhauser shift $\delta_n$ as a function
of laser energy for $\sigma^+$ ($\sigma^-$) laser polarization shown as solid circles (hollow triangles) for (A,B) X$^0$ of dot I, (C,D) X$^+$ of dot I, (E,F) X$^0$ of dot II and (G,H) X$^+$ of dot II.  
}\end{figure}

In the following we discuss the origin of the DNP that builds up through a simultaneous spin flip of an electron spin with a nuclear spin through the fluctuating term ($I_-^k S_+ + I_+^k S_-$) in equation 1 \cite{md2008,braun2006}.  This flip-flop process is repeated in time and a steady state nuclear polarization is reached. This process is very costly in energy for an electron in a neutral exciton X$^0$ \cite{bracker08}, as the bright and dark states (for example $\vert\Uparrow,\downarrow\rangle$ and  $\vert\Uparrow,\uparrow\rangle$) are separated in energy by the isotropic exchange energy of up to $\delta_0 \simeq 500 \mu$eV at zero magnetic field in InGaAs dots \cite{bu03}.  As a result the probability for electron-nuclear spin flip-flops is very low, so we argue that (i) the electron of the X$^0$ does not transfer its spin to the nuclei  (see below) (ii) X$^0$ is robust against decoherence by fluctuating nuclear spins which acts through the same flip-flop term. 
In the literature nuclear spin effects on carriers forming the X$^0$ have been studied in applied magnetic fields much larger than B$_{AEI}$ where the Zeeman effect dominates \cite{bracker08,GammonPRL}. As shown below, the X$^0$ is not at the origin of the DNP, but merely experiences the existing field B$_N$ in the dot. 

The PL in figure 1A shows that the dot is occupied by X$^0$ some of the time, by  X$^+$ for the rest of the time. The different origins of the charged excitons are discussed in the literature \cite{Nanotech}. In a simple picture, assuming that the dot contains a doping hole, the capture process for electrons (which are less likely to be trapped by potential fluctuations of the WL)  could be faster than for holes and an X$_0$ is formed. If no hole arrives within $t\le\tau_r$, the X$^0$ will recombine, if a hole is trapped for $t\le\tau_r$, the X$^+$ exciton is formed. Alternatively, a hole could tunnel into or out of the dot during $\tau_r$ to a nearby acceptor. During the radiative lifetime of the X$^+$ electron-nuclear spin flip-flop processes are far more likely as compared to the X$^0$ case, because in the absence of Coulomb exchange the energy difference between the X$^+$ states $\vert \Uparrow \Downarrow, \downarrow \rangle$ and $\vert \Uparrow \Downarrow , \uparrow \rangle$ is only $\simeq \delta_n$. As B$_N$ is essentially constant over at least ms \cite{maletinsky07},  the electron spin of the X$^0$ experiences the same B$_N$ as the electron in the X$^+$.

To demonstrate this point, we compare PLE measurements on the dot investigated already in figures 1 and 2 (dot I, see figure 3 upper part) with an additional dot II with a considerably larger splitting $\delta_1 = 34 \mu$eV (see figure 3 lower part). High values of P$_c$ in the order of 50\% are created for the X$^+$ in both dots when exciting with E$_\text{laser}$=1.425 to 1.465~eV. When approaching the low energy tail of the density of states of the WL at E$_\text{laser}\simeq$1.41~eV, the carrier absorption rate is too low to create nuclear polarization \cite{braun2006}. At E$_\text{laser} \geq$ 1.48~eV the P$_c$ drops in absolute value and even changes sign as the light hole transitions in the WL are excited \cite{Moskalenko02, Barrau93}. $\delta_n$ changes sign accordingly, which demonstrates that nuclear spin orientation in a QD can also be controlled via the optical selection rules in the WL. Comparing figures 3G and 3H for dot II shows clearly that the P$_c$ created for the X$^+$ is transferred to the nuclear spins. In stark contrast, the P$_c$ for the X$^0$ is on average zero in figure 3E. The neutral exciton X$^0$ in both dots is subject to a nuclear field of several hundred mT (figure 3B and 3F), created by the charged exciton state X$^+$, but for dot II B$_{AEI} \gg $ B$_{N}$, so the projection of the total effective magnetic field onto the \emph{z}-axis is too small to induce optical orientation. Although the values of $\delta_1$ differ by a factor of 3.5 due to different QD size, shape, composition and strain, the values of $\delta_0$ are less sensitive to the exact dot symmetry as they originate mainly from the short range Coulomb exchange \cite{Mb2,Oberli}. Similar values of $\delta_0 \gg \delta_1, \delta_n$  of typically a few hundred $\mu$eV can be assumed for both dots I and II. This means that the P$_c$ shown in figure 3A for dot I is due to the B$_N$ present in the dot, and not vice versa \cite{footBN}.

In summary, optical orientation of neutral excitons X$^0$ in single QDs in the absence of any applied fields is achieved as an effective nuclear magnetic field B$_N$  is constructed through non-resonant optical pumping. Varying B$_N$ in the presence of a constant B$_{AEI}$ due to Coulomb exchange allows efficient and tunable conversion of circularly to linearly polarized light mediated by a single QD.
Considering the slow evolution of B$_N$ \cite{maletinsky07} and the robustness of the electron spin during energy relaxation, our all optical approach could evolve in future experiments to orientate both the nuclear and the electron spins electrically in QD based Spin- Light Emitting Diodes \cite{Kawakami2001,Jonker08,Crowell06}. 

We thank ANR, IUF and DGA for financial support.


\begin{thebibliography}{36}
\bibitem[{foo({\natexlab{b}})}]{footLH}
\bibinfo{note}{Light hole levels in the dot are several tens of meV away in
  energy due to strain and are not considered here.}
\bibitem[{foo({\natexlab{c}})}]{foot1}
\bibinfo{note}{Although resonant PL experiments show that a linearly polarized
  excitation does result in linearly polarized PL, the coherent superposition
  between $\vert\Uparrow,\downarrow\rangle$ and
  $\vert\Downarrow,\uparrow\rangle$ created in non-resonant excitation is lost
  during carrier capture. The spin information is solely carried by the
  electron and we have found no correlation between the linear polarization of
  the laser and P$_c$ or P$_{l}$ of the PL.}
\bibitem[{foo({\natexlab{d}})}]{footFIT}
\bibinfo{note}{A single Lorentzian is fitted to a doublet for which the energy
  splitting is much smaller than it's linewidth. It can be shown that changes
  in the oscillator strength of $ \vert + \rangle$ and $\vert - \rangle$ (i.e.
  changes in $\alpha$ and $\beta$) shift this single Lorentzian by an energy
  $\delta_n$.}
\end{thebibliography}

\end{document}